\documentclass[12pt,a4paper,twoside]{article}
\usepackage{amsmath,amssymb,amscd}
\numberwithin{equation}{section} 

\newtheorem{theorem}{Theorem}[section]
\newtheorem{lemma}[theorem]{Lemma}
\newtheorem{definition}[theorem]{Definition}
\newtheorem{remark}[theorem]{Remark}

\newtheorem{proposition}[theorem]{Proposition}

\newenvironment{proof}{\paragraph{Proof.}}{\hfill $\square$\\}
\newenvironment{proof*}{\paragraph{Proof.}}{}

\newcommand{\hk}{\hslash}
\newcommand{\arrow}{\rightarrow}

\newcommand{\diff}[2]{\frac{d#1}{d#2}}

\newcommand{\Z}{\mathbb{Z}}

\newcommand{\Rm}{\mathbb{R}}
\newcommand{\Km}{\mathbb{K}}

\newcommand{\Time}{\mathbb{T}}
\newcommand{\G}{\mathcal{G}}

\newcommand{\pr}{\partial}

\newcommand{\bra}[1]{\left (#1\right )}
\newcommand{\brac}[1]{\left [#1\right ]}

\newcommand{\e}{\mathcal{E}}
\newcommand{\pmatrx}[1]{\begin{pmatrix} #1 \end{pmatrix}}

\newcommand{\K}{{\mathbb K}}

\begin{document}

\title{$R$-matrix approach to integrable systems on time scales}

\author{Maciej B\l aszak$^\dag$, Burcu Silindir$^\ddag$ and B\l a\.zej M. Szablikowski$^\dag$\\[3mm]
\small (\dag) Department of Physics, Adam Mickiewicz University\\
\small Umultowska 85, 61-614 Pozna\'n, Poland\\
\small e-mail's: {\tt blaszakm@amu.edu.pl} and {\tt bszablik@amu.edu.pl}\\[2mm]
\small (\ddag) Department of Mathematics, Faculty of Sciences\\
\small Bilkent University, 06800 Ankara, Turkey\\
\small e-mail: {\tt silindir@fen.bilkent.edu.tr}}
\date{}
\maketitle

\begin{abstract}
A general unifying framework for integrable soliton-like systems
on time scales is introduced. The $R$-matrix formalism is applied
to the algebra of $\delta$-differential operators in terms of which
one can construct infinite hierarchy of commuting vector fields.
The theory is illustrated by two infinite-field integrable
hierarchies on time scales which are difference counterparts of KP
and mKP. The difference counterparts of AKNS and Kaup-Broer
soliton systems are constructed as related finite-field
restrictions.
\end{abstract}

\section{Introduction}

Integrable systems are widely investigated in $(1+1)$ dimensions,
where one of the dimensions stands for the time evolution variable
and the other one stands for the space variable. The space
variable is usually considered on continuous intervals, or both on
integer values and on $\mathbb{R}$ \cite{MB} or on $\mathbb{K}_q$
intervals \cite{frenkel,Adler}. In order to embed the study of
integrable systems into a more general unifying framework, one of
the possible approaches is to construct the integrable systems on
time scales. Here the space variable is considered on any time
scale where $\mathbb{R}$, $\hk\Z$, $\mathbb{K}_q$ are special
cases. The first step in this direction was taken in \cite{G-G-S}, where the Gelfand-Dickey approach
\cite{gelfand,bookdickey} was extended in order to construct
integrable nonlinear evolutionary equations on any time scale.
Another unifying approach is to formulate different types of
discrete dynamics on $\mathbb{R}$. Some contribution in this
direction was made recently in \cite{bgss}.

The main goal of this work is to present a theory for the systematic
construction of $(1+1)$-dimensional integrable systems on time
scales in the frame of the $R$-matrix formalism. By an integrable
system, we mean such a system which has an infinite-hierarchy of
mutually commuting symmetries. The $R$-matrix formalism is one of the
most effective and systematic methods of constructing integrable
systems \cite{reyman,semenov}. This formalism originated from
 the pioneering article \cite{gelfand} by Gelfand and Dickey,
 who constructed the soliton systems of KdV type. The crucial point
of the $R$-matrix formalism is that the construction of integrable
systems proceeds from the Lax equations on appropriate Lie
algebras  \cite{reyman,semenov}.  The simplest $R$-matrices can be
constructed by a decomposition of a given  Lie algebra into two
Lie subalgebras. We refer to \cite{semenov,bookdickey,MB} for
abstract formalism of classical $R$-matrices on Lie algebras.

This paper is organized as follows: In the next section, we give a
brief review of the time scale calculus. In the third section, we
define the $\delta$-differentiation operator and formulate the Leibniz
rule for this operator. We introduce the Lie algebra as an algebra
of $\delta$-differential operators equipped with the commutator,
decompose it into two Lie subalgebras and construct the simplest
$R$-matrix on this algebra. We present  the appropriate Lax
operators for infinite-field cases and the admissible finite-field
restrictions generating consistent Lax hierarchies. In $\Time=\Rm$
case, or in the continuous limit of some special  time scales, we
observe that the algebra of $\delta$-differential operators turns
out to be the algebra of pseudo-differential operators. Next, we
formulate and prove the property of the algebra of $\delta$-differential
operators. This property allows us to obtain natural constraints
which are fulfilled by finite field restrictions. Therefore,  the
source of the constraints, obtained in the Burgers equations and KdV
hierarchy on time scales in \cite{G-G-S}, is established. We end
up this section with the construction of the recursion operators
by means of the method presented in \cite{Gurses}. In the fourth
section, we illustrate two infinite-field integrable hierarchies
on time scales which are difference counterparts of
Kadomtsev-Petviashvili (KP) and modified Kadomtsev-Petviashvili
(mKP) hierarchies. In the last section, we present finite-field
restrictions which are difference counterparts of
Ablowitz-Kaup-Newell-Segur (AKNS) and Kaup-Broer (KB) hierarchies
with  their recursion operators.

\section{Preliminaries}

In this section, we give a brief introduction to the concept of
time scale. We refer to \cite{boh1,boh2} for the basic definitions
and  general theory of time scale. What we mean by a time scale
$\Time$, is an arbitrary nonempty closed subset of real numbers.
The time scale calculus was introduced by Aulbach and Hilger
\cite{ah,hil} in order to unify all possible intervals on the real
line $\mathbb{R}$, like continuous (whole) $\mathbb{R}$, discrete
$\mathbb{Z}$, and $q$-discrete $\mathbb{K}_q$ $({\mathbb
K}_{q}=\,q^{\mathbb Z}\cup \{0\} \equiv \{ q^{k}: k \in {\mathbb
Z}\} \cup \{0\}$, where $q\neq 1$ is a fixed real number)
intervals. For the definition of the derivative in time scales, we use
\textit{forward} and \textit{backward jump operators} which are
defined as follows.

\begin{definition}
For $x \in \Time$, the forward jump operator $\sigma: {\mathbb
T}\rightarrow {\mathbb T}$ is defined by
\begin{equation}
\sigma(x)=\inf\, \{ y \in {\mathbb T}: y> x\},
\end{equation}
while the backward jump operator $\rho: {\mathbb T} \rightarrow
{\mathbb T}$ is defined by
\begin{equation}
\rho(x)=\sup\, \{ y \in {\mathbb T}: y < x\}.
\end{equation}
We set in addition $\sigma(\max\Time) = \max \Time$ if there
exists a finite $\max\Time$, and  $\rho(\min\Time) = \min \Time$
if there exists a finite $\min\Time$.

The jump operators $\sigma$  and $\rho$ allow the classification
of points in a time scale in the following way:  $x$ is called right
dense, right scattered, left dense, left scattered, dense and
isolated if $\sigma(x)=x, \ \sigma(x)>x, \ \rho(x)=x, \ \rho(x)<x,
\ \sigma(x)=\rho(x)=x$ and $\rho(x)<x<\sigma(x)$, respectively.
Moreover, we define  the graininess functions $\mu,\, \nu :
{\mathbb T} \rightarrow [0,\infty)$ as follows
\begin{equation}
\mu(x)=\sigma (x)-x, \quad\nu(x)=x-\rho(x),\quad \mbox{for all}~ x
\in {\mathbb T}.
\end{equation}\end{definition}

In literature, ${\mathbb T}^{\kappa}$ denotes a set
consisting of ${\mathbb T}$ except for a possible left-scattered
maximal point while  ${\mathbb T}_{\kappa}$ stands for a set of
points of ${\mathbb T}$ except for  a possible right-scattered
minimal point.

\begin{definition}\label{derivative}
Let $f:\mathbb{T} \to \mathbb{R}$ be a function on a time scale
$\Time$. For $x\in\mathbb{T}^\kappa$, delta derivative of $f$,
denoted by $\Delta f$, is defined as
\begin{equation}\label{del}
\Delta f(x) = \lim_{s\to x} \frac{f(\sigma (x))-f(s)}{\sigma
(x)-s},\qquad s\in\Time,
\end{equation}
while for $x\in \mathbb{T}_\kappa$, $\nabla$-derivative of $f$,
denoted by $\nabla f$, is defined as
\begin{equation}\label{nab}
\nabla f(x) = \lim_{s\to x} \frac{f(s)-f(\rho(x))}{s-\rho
(x)},\qquad s\in\Time,
\end{equation}
provided that the limits exist. A function $f:{\mathbb T}
\rightarrow {\mathbb R}$ is said to be $\Delta$-smooth
($\nabla$-smooth) if it is infinitely $\Delta$-differentiable
($\nabla$-differentiable).
\end{definition}
\begin{remark}Let $f:\mathbb{T}\to\mathbb{R}$ be $\Delta$-differentiable on
$\mathbb{T}^\kappa$. If $x$ is right-scattered, then the
definition \eqref{del} turns out to be
\begin{equation*} \Delta f(x)= \frac{f(\sigma(x))-f(x)}{\mu(x)},
\end{equation*}
while if $x$ is right-dense, \eqref{del} implies that
\begin{equation*}
\Delta f(x)=\lim_{s \rightarrow x}\, \frac{f(x)-f(s)}{x-s},\qquad
s\in\Time.
\end{equation*}
Similarly, let $f:\mathbb{T}\to\mathbb{R}$ be
$\nabla$-differentiable on $\mathbb{T}_\kappa$. If $x$ is
left-scattered, then the definition \eqref{nab} turns out to be
\begin{equation*}
\nabla f(x)= \frac{f(x)-f(\rho(x))}{\nu(x)},
\end{equation*} while if $x$ is left-dense, \eqref{nab} yields as
\begin{equation*}
\nabla f(x)=\lim_{s \rightarrow x}\, \frac{f(x)-f(s)}{x-s},\qquad
s\in\Time.
\end{equation*}
\end{remark}

In order to be more precise, we present $\Delta$ and $\nabla$
derivatives for some special time scales. If ${\mathbb T}={\mathbb
R}$, then $\Delta$- and $\nabla$-derivatives become ordinary
derivatives, i.e.
\begin{equation*}
    \Delta f(x) = \nabla f(x) = \frac{df(x)}{dx}.
\end{equation*}
If ${\mathbb T}=\hk\Z$, then
\begin{equation*}
\Delta f(x)= \frac{f(x+\hk)-f(x)}{\hk}~~~~ \mbox{and}~~~~\nabla
f(x)= \frac{f(x)-f(x-\hk)}{\hk}.
\end{equation*}
If ${\mathbb T}={\mathbb K}_{q}$, then
\begin{equation*}
\Delta f(x)= \frac{f(qx)-f(x)}{(q-1)x} ~~~\mbox{and}~~~\nabla
f(x)= \frac{f(x)-f(q^{-1}\,x)}{(1-q^{-1})x},
\end{equation*}
for all $x \ne 0$, and
\begin{equation*}
\Delta f(0)=\nabla f(0)=\lim_{s \rightarrow 0} \frac{f(s)-f(0)}{s},\quad s\in\mathbb{K}_q,
\end{equation*}
provided that this limit exists.

As an important property of $\Delta$-differentiation on $\mathbb
T$, we give the product rule. If $f,g:\mathbb{T} \to \mathbb{R}$
are $\Delta$-differentiable functions at $x\in \mathbb
T^{\kappa}$, then their product is also $\Delta$-differentiable
and the following Lebniz-like rule hold
\begin{equation}\label{leib}
   \begin{split}
   \Delta(f g)(x) &= g(x)\Delta f(x) + f(\sigma(x))\Delta g(x)\\
                  &= f(x)\Delta g(x) + g(\sigma(x))\Delta f(x).
   \end{split}
\end{equation}
Besides, if $f$ is $\Delta$-smooth function, then
\begin{equation}\label{rel}
    f(\sigma(x)) = f(x) + \mu(x)\Delta f(x).
\end{equation}
If $x\in\Time$ is right-dense, then $\mu(x)=0$ and the relation
\eqref{rel} is trivial.

\begin{definition}
A time scale ${\mathbb T}$ is regular if both of the following two
conditions are satisfied:
\begin{itemize}
  \item[(i)]  $\sigma(\rho(x))=x$ for all $x\in\Time$,
  \item[(ii)] $\rho(\sigma(x))=x$ for all $x\in\Time$.
\end{itemize}
\end{definition}

Set $x_*=\min\Time$ if there exists a finite $\min\Time$, and set
$x_* = -\infty$ otherwise. Also set $x^*=\max\Time$ if there
exists a finite $\max\Time$, and set $x^* = \infty$ otherwise.

\begin{proposition}\emph{\cite{G-G-S}}
A time scale is regular if and only if the following two
conditions hold:
\begin{itemize}
  \item[(i)] the point $x_*=\min\Time$ is right dense and the point $x^*=\max\Time$
  is left-dense;
  \item[(ii)] each point of $\Time\setminus \{x_*,x^*\}$ is either two-sided
  dense or two-sided scattered.
\end{itemize}
\end{proposition}

In particular $\mathbb{R}, \hk\mathbb{Z}$ ($\hk\neq 0$) and
$\mathbb{K}_q$ are regular time scales, as are $[0,1]$ and $[-1,0]
\cup \{1/k:k\in\mathbb{N}\} \cup \{k/(k+1):k\in\mathbb{N}\} \cup
[1,2]$.

Throughout this work, let $\Time$ be a regular time scale. By
$\Delta$, we denote the delta-differentiation operator which
assigns  each $\Delta$-differentiable function
$f:\mathbb{T}\to\mathbb{R}$  to its delta-derivative $\Delta(f)$,
defined by
\begin{equation}
[\Delta(f)](x)=\Delta f(x), \quad\mbox{for}\quad x \in {\mathbb
T}^{\kappa}.
\end{equation}
The {\it shift operator} $E$ is defined by the formula
\begin{equation}
(Ef)(x)=f(\sigma(x)),\qquad x \in {\mathbb T}.
\end{equation}
 The inverse $E^{-1}$ is defined by
\begin{equation}
(E^{-1}\,f)(x)=f(\sigma^{-1}(x))=f(\rho(x)),
\end{equation}
for all $x\in\Time$. Note that $E^{-1}$ exists only in the case of
regular time scales and that in general $E$ and $E^{-1}$ do not
commute with $\Delta$ and $\nabla$ operators.

\begin{proposition}\emph{\cite{aticiguseinov}}
Let $\mathbb T$ be a regular time scale.
\begin{itemize}
\item[(i)] If $f: {\mathbb T} \rightarrow {\mathbb R}$ is a
$\Delta$-smooth function on $\mathbb T^{\kappa}$, then $f$ is
$\nabla$-smooth and for all $x\in \mathbb{T}_{\kappa}$,
\begin{equation}\label{nabladelta}
 \nabla f(x)= E^{-1}\Delta f(x).
\end{equation}
\item[(ii)] If $f: {\mathbb T} \rightarrow {\mathbb R}$ is a
$\nabla$-smooth function on $\mathbb T_{\kappa}$, then $f$ is
$\Delta$-smooth and for all $x\in \mathbb{T}^{\kappa}$,
\begin{equation}\label{deltanabla}
 \Delta f(x)= E\nabla f(x).
\end{equation}
\end{itemize}
\end{proposition}

Thus the properties of $\Delta$- and $\nabla$-smoothness for
functions on regular time scales are equivalent.

In some special cases, by properly introducing the deformation
parameter, it is possible to consider a continuous limit of a time
scale. For instance, the continuous limit of $\hk\Z$ is the whole
real line $\Rm$, i.e.
\begin{equation}\label{c1}
\begin{CD}
    \Time = \hk\Z @>\hk\arrow 0>> \Time=\Rm;
\end{CD}
\end{equation}
and the continuous limit of $\K_q$ is the closed half line
$\Rm_{+}\cup{0}$, thus
\begin{equation}\label{c2}
\begin{CD}
    \Time = \K_q @>q\arrow 1>> \Time=\Rm_{+}\cup{0}.
\end{CD}
\end{equation}

For more about the calculus on time scales we refer the readers to
\cite{boh1,boh2}.

\section{Algebra of $\delta$-differential operators}

\subsection{Basic notions}

In this section, we deal with the algebra of $\delta$-differential
operators defined on a regular time scale $\Time$. We denote the
delta differentiation operator by $\delta$ instead of $\Delta$,
for convenience in the operational relations. The operator $\delta
f$ which is a composition of $\delta$ and $f$, where $f:
\Time\rightarrow {\mathbb R}$, is introduced as follows
\begin{equation}\label{delta}
\delta f:=\Delta f+E(f) \delta, \quad \forall f.
\end{equation}
Note that, the definition \eqref{delta} is consistent with the
Lebniz-like rule on time scales \eqref{leib}.

\begin{theorem}
The Leibniz rule on time scales for the operator $\delta$ is given
as follows.
\begin{itemize}
  \item[(i)] For $n\geqslant 0$:
\begin{equation}
\delta^{n}
f=\sum_{k=0}^n\quad\sum_{i_1+i_2+...+i_{k+1}=n-k}(\Delta^{i_{k+1}}E\Delta^{i_{k}}E...\Delta^{i_{2}}E\Delta^{i_{1}})
f\delta^k,\label{leibnizpositive}
\end{equation}
where $i_\gamma \geqslant 0$ for all $\gamma=1,2,..,k+1$. Here the
formula includes all possible strings containing $n-k$ times
$\Delta$ and $k$ times $E$.
  \item[(ii)] For $n<0$:
\begin{equation}
\delta^{n}
f=\sum_{k=-n}^\infty\quad\sum_{i_1+i_2+...+i_{k+n+1}=k}(-1)^{k+n}(E^{-i_{k+n+1}}\Delta
E^{-i_{k+n}}\Delta...E^{-i_{2}}\Delta E^{-i_{1}})
f\delta^{-k},\label{leibniznegative}
\end{equation}
where $i_\gamma > 0$ for all $\gamma=1,2,..,k+n+1>0$. Here the
formula includes all possible strings containing $k+n+1$ times $E$
and $k+n$ times $\Delta$.
\end{itemize}
\end{theorem}

The above theorem is a straightforward consequence of definition
\eqref{delta}. Note that $\delta^{-1} f$ has the form of the
formal series
\begin{equation}
\delta^{-1}f=\sum_{k=0}^{\infty}(-1)^{k}((E^{-1} \Delta)^k
E^{-1})f\delta^{-k-1},\label{leibniz2}
\end{equation}
which was previously  given in \cite{G-G-S}, in terms of $\nabla$.
Thus \eqref{leibniznegative} is the appropriate generalization of
\eqref{leibniz2}.

\subsection{Classical $R$-matrix formalism}

In order to construct integrable hierarchies of mutually commuting
 vector fields on time scales, we deal  with a systematic method, so-called {\it the classical
 $R$-matrix formalism} \cite{semenov,bookdickey,MB}, presented in the following
scheme.

Let $\mathcal{G}$ be an algebra, with some associative
multiplication operation, over a commutative field $\mathbb{K}$ of
complex or real numbers, based on an additional bilinear product
given by a Lie bracket $[\cdot,\cdot]:\mathcal{G}\to\mathcal{G}$,
which is skew-symmetric and satisfies the Jacobi identity.

\begin{definition}\label{Rmatrix}
A linear map $R:\mathcal{G}\to\mathcal{G}$ such that the bracket
\begin{equation}
[a,b]_R:=[Ra,b]+[a,Rb],\label{rmatrix}
\end{equation}
is a second Lie bracket on $\mathcal{G}$, is called the classical
$R$-matrix.
\end{definition}

Skew-symmetry of \eqref{rmatrix} is obvious. When one  checks the
Jacobi identity of \eqref{rmatrix}, it can be clearly deduced that
 a sufficient condition for $R$ to be a classical $R$-matrix is
\begin{equation}\label{yangbaxter}
[Ra,Rb]-R[a,b]_R+\alpha[a,b]=0,
\end{equation}
where $\alpha\in\mathbb{K}$, called the \textit{Yang-Baxter
equation} YB$(\alpha)$. There are only two relevant cases of
YB$(\alpha)$, namely $\alpha\neq 0$ and $\alpha=0$, as Yang-Baxter
equations for $\alpha\neq 0$ are equivalent and can be
reparametrized.

Additionally, assume that the Lie bracket is a derivation of
multiplication in $\G$, i.e. the relation
\begin{equation}\label{Rmatrixleibniz}
[a,bc]=b[a,c]+[a,b]c\qquad a,b,c\in\G
\end{equation}
holds. If the Lie bracket is given by the commutator, i.e. $[a,b]=
ab-bc$, the condition \eqref{Rmatrixleibniz} is satisfied
automatically, since $\G$ is associative.

\begin{proposition}\label{proposition}
Let $\G$ be a Lie algebra fulfilling all the above assumptions and
$R$ be the classical $R$-matrix satisfying the Yang-Baxter
equation, YB$(\alpha)$. Then the power functions $L^n$ on $\G$,
$L\in\G$ and $n\in\mathbb{Z}_+$, generate the so-called Lax
hierarchy
\begin{equation}\label{vectorfield}
\frac{dL}{dt_n} = \brac{R(L^n),L},
\end{equation}
of pairwise commuting vector fields on $\G$. Here, $t_n$'s are
related evolution parameters. We additionally assume that $R$
commutes with derivatives with respect to these evolution
parameters.
\end{proposition}
\begin{proof}
It is clear that the power functions on $\G$ are well defined.
 Then
\begin{align*}
    (L_{t_m})_{t_n}-(L_{t_n})_{t_m} &= [RL^m,L]_{t_n}-[RL^n,L]_{t_m}\\
&=[(RL^m)_{t_n}-(RL^n)_{t_m},L]+[RL^m,[RL^n,L]]-[RL^n,[RL^m,L]]\\
&=[(RL^m)_{t_n}-(RL^n)_{t_m}+[RL^m,RL^n],L].
\end{align*}
Hence, the vector fields \eqref{vectorfield} mutually commute if
the so-called {\it zero-curvature} (or Zakharov-Shabat) {\it
equations}
\begin{equation*}
(RL^m)_{t_n}-(RL^n)_{t_m}+[RL^m,RL^n]=0,
\end{equation*}
are satisfied. From \eqref{vectorfield} and by the Leibniz rule
\eqref{Rmatrixleibniz} we have that $(L^m)_{t_n} = [RL^n,L^m]$.
Using Yang-Baxter equation for $R$ and the fact that $R$ commutes
with $\pr_{t_n}$, we deduce
\begin{align*}
R(L^m)_{t_n} &- R(L^n)_{t_m} + [RL^m,RL^n] =\\
&= R[RL^n, L^m] - R[RL^m,L^n] + [RL^m,RL^n]\\
&= [RL^m,RL^n]-R[L^m,L^n]_R = - \alpha [L^m,L^n] = 0.
\end{align*}
Hence, the vector fields pairwise commute.
\end{proof}

In practice the powers of Lax operators in \eqref{vectorfield} are
fractional. Notice that, the Yang-Baxter equation is a sufficient
condition for mutual commutation of vector fields
\eqref{vectorfield}, but not necessary. Thus choosing an algebra
$\G$ properly, the Lax hierarchy yields abstract integrable
systems. In practice, the element $L$ of $\G$ must be
appropriately chosen, in such a way that the evolution systems
\eqref{vectorfield} are consistent on the subspace of $\G$.

\subsection{Classical $R$-matrix on time-scales}

We introduce the algebra $\mathcal{G}$ as an algebra of formal
Laurent series of (pseudo-) $\delta$-differential operators
equipped with the commutator, and define its decomposition such as:
\begin{equation}\label{algebra}
 \G= \G_{\geqslant k}\oplus \G_{< k}= \{ \sum_{i\geqslant k}u_{i}(x)\delta^{i}\}\oplus
 \{\sum_{i<k}u_{i}(x)\delta^{i}\},
\end{equation}
where $u_{i}:\Time\arrow \Km$ are $\Delta$-smooth functions. The
subspaces $\mathcal{G}_{\geqslant k}$, $\mathcal{G}_{< k}$ are
closed Lie subalgebras of $\mathcal{G}$ only if $k=0,1$. Thus, we
define the classical $R$-matrix in the following form
\begin{equation}\label{classicalrmatrix}
R:=\frac{1}{2}(P_{\geqslant k}-P_{<k})\qquad k=0,1,
\end{equation}
where $P_{\geqslant k}$ and $P_{<k}$ are the projections onto
$\G_{\geqslant k}$ and $\G_{<k}$, respectively. Since the
classical $R$-matrices \eqref{classicalrmatrix} are defined
 through the projections onto Lie subalgebras, they satisfy the Yang-Baxter equation
\eqref{yangbaxter} for $\alpha=\frac{1}{4}$.

Let $L\in\mathcal{G}$ be given in the form
\begin{equation}\label{laxo}
L=u_N\delta^N+u_{N-1}\delta^{N-1}+\ldots+u_1\delta+u_0+u_{-1}\delta^{-1}+\ldots,
\end{equation}
where $u_i$ are dynamical fields depending additionally on
the evolution parameters $t_n$. Thus, the Lax hierarchy
\eqref{vectorfield}, based on \eqref{classicalrmatrix} and in
general generated  by fractional powers of $L$, turns out to be
\begin{equation}\label{laxequation}
\frac{dL}{dt_n}=\brac{\bra{L^\frac{n}{N}}_{\geqslant
k},L}=-\brac{\bra{L^\frac{n}{N}}_{<k},L}\qquad k=0,1\qquad
n\in\mathbb{Z}_{+}.
\end{equation}
 Proposition \ref{proposition} implies that the hierarchy
\eqref{laxequation} is  infinite hierarchy of mutually commuting
vector fields and represents $(1+1)$-dimensional integrable
differential-difference systems on a time scale $\Time$, including the
time variables $t_n$ and space variable $x\in\Time$.

Analyzing \eqref{laxequation} for $L$ given by \eqref{laxo}, in
the case of $k=0$, one finds that $(u_N)_t=0$ and $(u_{N-1})_t =
\mu(\ldots)$ (see also Remark \ref{remark1}). Similarly for $k=1$,
we have $(u_N)_t= \mu(\ldots)$ (see also Remark \ref{remark2}).
Hence, the appropriate Lax operators, yielding consistent Lax
hierarchies \eqref{laxequation}, are in the following form:
\begin{align}
\label{l1} k=0:\qquad & L=c_N\delta^N+\tilde{u}_{N-1}\delta^{N-1}+\ldots+u_1\delta^1+u_0+u_{-1}\delta^{-1}+\ldots\\
\label{l2} k=1:\qquad &
L=\tilde{u}_N\delta^N+u_{N-1}\delta^{N-1}+\ldots+u_1\delta^1+u_0+u_{-1}\delta^{-1}+\ldots,
\end{align}
where $c_N$ is a time-independent field and fields
$\tilde{u}_{N-1},\tilde{u}_N$ are time-independent for dense
$x\in\Time$, as at these points $\mu = 0$. This is the reason why
they are distinguished by a tylde mark.

Nevertheless, we are interested in finite-field integrable systems
on time-scales. Thus, in order to work with a finite number of
fields, we should impose some restrictions on \eqref{l1} and
\eqref{l2} in such a way that the commutator on the right-hand side of
the Lax equation \eqref{laxequation} does not produce terms not
contained in the left-hand side of the Lax equation. To be more
precise, the left- and right-hand of \eqref{laxequation} span the
same subspace of $\mathcal{G}$. From this purpose, in the case of
$k=0$, one finds the general admissible form of finite-field Lax
operator given by
\begin{equation}\label{finiterestriction0}
    L = c_N\delta^N+\tilde{u}_{N-1}\delta^{N-1}+\ldots+u_1\delta+u_0+\sum_s\psi_s\delta^{-1}\varphi_s,
\end{equation}
with further restriction
\begin{equation}\label{finiterestriction01}
  L = c_N\delta^N+\tilde{u}_{N-1}\delta^{N-1}+\ldots+u_1\delta+u_0.
\end{equation}
 In the case of $k=1$, the general admissible Lax operator has the form
\begin{equation}\label{finiterestriction1}
    L = \tilde{u}_N\delta^N+u_{N-1}\delta^{N-1}+\ldots+u_1\delta+u_0+\delta^{-1}u_{-1}
    +\sum_s\psi_s\delta^{-1}\varphi_s,
\end{equation}
and further restrictions are
\begin{align}
\label{finiterestriction10} L &= \tilde{u}_N\delta^N+u_{N-1}\delta^{N-1}+\ldots+u_1\delta+u_0+\delta^{-1}u_{-1}\\
\label{finiterestriction11} L &= \tilde{u}_N\delta^N+u_{N-1}\delta^{N-1}+\ldots+u_1\delta+u_0\\
\label{finiterestriction12} L &=
\tilde{u}_N\delta^N+u_{N-1}\delta^{N-1}+\ldots+u_1\delta.
\end{align}
In the above Lax operators $c_N$ is a time-independent field for all
$x\in\Time$ and $\tilde{u}_{N-1}, \tilde{u}_N$ are
time-independent at dense points from a time scale. We assume also
that  the sum $\sum_s$ is finite.

In general, for an arbitrary regular time scale $\Time$, the Lax
hierarchies \eqref{laxequation} represent hierarchies of
soliton-like integrable difference systems. For instance, when
$\Time = \hk\Z$ or $\K_q$, the hierarchies \eqref{laxequation} are
those of lattice and $q$-deformed (-like) (discrete) soliton
systems, respectively. In particular, for the case of $\Time=\Rm$,
i.e. the continuous time scale on the whole $\Rm$, the Lax
hierarchies are those of field soliton systems. In some
cases, field soliton systems can also be obtained from the
continuous limit of integrable systems on time scales (see
\eqref{c1} and \eqref{c2}).

In the continuous time scale, the algebra of $\delta$-differential
operators \eqref{algebra} turns out to be the algebra of
pseudo-differential operators
\begin{equation}\label{algebrapartial}
 \mathcal{G}=\mathcal{G}_{\geqslant k}\oplus \mathcal{G}_{< k}= \{ \sum_{i\geqslant
 k}
u_{i}(x)\partial^{i}\}\oplus \{ \sum_{i< k}
u_{i}(x)\partial^{i}\},
\end{equation}
where $\pr$ is such that $\pr u = \pr_x u + u\pr=u_x+u\pr$. The
above decomposition is valid only if $k=0,1$ and $2$. Thus, in the
general theory of integrable systems on time scales, we loose one
case in contrast to the ordinary soliton systems constructed by means of
pseudo-differential operators. This follows from the fact that,
for $k=2$, \eqref{algebra} does not decompose into Lie subalgebras
for an arbitrary time scale. For appropriate Lax operators, finite
field restrictions and more information about the algebra of
pseudo-differential operators, we refer the reader to
\cite{oevel,oevelsrt,bookdickey,MB}. Note that the fields $\psi_s$
and $\varphi_s$ in \eqref{finiterestriction0} and
\eqref{finiterestriction1} are special dynamical fields in the case of
the algebra of pseudo-differential operators. They are the 
so-called source terms, as  $\psi_s$ and $\varphi_s$ are
eigenfunctions and adjoint-eigenfunctions, respectively, of the
Lax hierarchy \eqref{laxequation} \cite{oevelsrt}.

It turns out that there are constraints between dynamical fields
of the admissible finite-field Lax restrictions
\eqref{finiterestriction0}-\eqref{finiterestriction12} fulfilling
\eqref{laxequation}. We give these constraints in the following
theorem, which is a consequence of the property of the algebra of
$\delta$-differential operators. This property is illustrated in
the following lemma.

\begin{lemma}\label{lemma1}
Consider the equality
\begin{equation}\label{l1r}
    \delta^r F=\sum_{i=0}^{r}C_i\delta^{r-i},\qquad r>0.
\end{equation}
Then the following relation
\begin{equation}\label{lemma1result}
\sum_{i=0}^r(-\mu)^iC_i=F
\end{equation}
is valid.
\end{lemma}
\begin{proof}
We make use of induction. Assume that \eqref{lemma1result} holds
for $r$. Then
\begin{align}
 \delta^{r+1}F = \delta^r(EF)\delta + \delta^r\Delta F = \sum_{i=0}^r A_i \delta^{r-i+1} +\sum_{i=0}^r B_i \delta^{r-i} = \sum_{i=0}^{r+1} C_i \delta^{r+1-i}.
\end{align}
By the assumption we have  $\sum_{i=0}^r(-\mu)^iA_i = EF$ and
$\sum_{i=0}^r(-\mu)^iB_i = \Delta F$. Hence
\begin{align}
  \sum_{i=0}^{r+1}(-\mu)^iC_i =  \sum_{i=0}^r(-\mu)^{(i+1)}B_i +  \sum_{i=0}^r(-\mu)^iA_i
        = -\mu\Delta F + EF = F.
\end{align}
\end{proof}

Let us explain the source of Lemma \ref{lemma1}. Consider the
equality
\begin{equation}\label{rr}
    A = \sum_{i\geqslant 0}a_i\delta^i =0,
\end{equation}
where the sum is finite, and $A$ is purely $\delta$-differential
operator. We expand $A$ with respect to the shift operator $\e$:
$\e u = E(u) \e$. From the relation \eqref{rel} we have
\begin{equation}\label{rell}
    \e = 1 + \mu\delta.
\end{equation}
The equality from Lemma \ref{lemma1} is trivially satisfied for
dense $x\in\Time$, since in this case $\mu=0$. Thus, it is enough
to consider remaining points in a time scale so assume that
$\mu\neq 0$. Hence, from \eqref{rell}, we have the formula
\begin{equation}
    \delta = {\mu}^{-1}\e - {\mu}^{-1}\label{abc}.
\end{equation}
Thus, using  \eqref{abc} the relation \eqref{rr}  can be rewritten
as
\begin{equation}
    A = \sum_{i}a'_i\e^i =0.
\end{equation}
Obviously, it must hold for terms of all orders. The equality for
the zero-order terms, i.e. $a'_0=0$, can be simply obtained by
replacing $\delta$ with $-{\mu}^{-1}$ in \eqref{rr}. The same
substitution in \eqref{l1r} allows us to find
\begin{equation}
    (-\mu)^{-r} F=\sum_{i=0}^{r}C_i (-\mu)^{-r+i},
\end{equation}
which is equivalent to \eqref{lemma1result}.

The above procedure can be extended also to operators $A$ that are
not purely $\delta$-differential and contain finitely many terms
with $\delta^{-1},\delta^{-2},\ldots$. As an illustration consider
the equality
\begin{equation}
[A\delta^r,\psi\delta^{-1}\varphi] =
\sum_{i=0}^{r-1}C_i\delta^{r-1-i} + \hat{C}_r\delta^{-1}\varphi +
\psi\delta^{-1}C_r.
\end{equation}
The above equality is well-formulated  since it follows
immediately from the definition and the property of the $\delta$
operator. Replacing $\delta$ with $-\mu^{-1}$, the commutator
vanishes, and we have
\begin{align}
   &0 = \sum_{i=0}^{r-1}C_i(-\mu)^{-r+1+i} + \hat{C}_r(-\mu)\varphi + \psi(-\mu)C_r
   \quad\Longleftrightarrow\\
    &\sum_{i=0}^{r-1}(-\mu)^iC_i + (-\mu)^r(\hat{C}_r\varphi + \psi C_r) = 0.
\end{align}
Straightforward consequence of such a behavior of
$\delta$-differential operators is the next theorem.

\begin{theorem}\label{theorem2}\hfill
\begin{itemize}
  \item[(i)] The case $k=0$. The constraint between dynamical fields of \eqref{finiterestriction0}, generating Lax hierarchy \eqref{laxequation},
  has the form
      \begin{equation}\label{con1}
        \begin{split}
      &(-\mu)^{N-1}\diff{\tilde{u}_{N-1}}{t_n} + \sum_{i=0}^{N-2}(-\mu)^i\diff{u_i}{t_n} - \mu\sum_s\diff{(\psi_s\varphi_s)}{t_n} = 0\\
      &\qquad\Longrightarrow\qquad
       (-\mu)^{N-1}\tilde{u}_{N-1} + \sum_{i=0}^{N-2}(-\mu)^i u_i - \mu\sum_s \psi_s\varphi_s = a_n,
        \end{split}
      \end{equation}
  where $n\in\Z_+$ and $a_n$ is a time-independent function.
  \item[(ii)] The case $k=1$. The constraint between dynamical fields of \eqref{finiterestriction1}, generating \eqref{laxequation}, has the form
      \begin{equation}\label{con2}
        \begin{split}
          & (-\mu)^{N}\diff{\tilde{u}_{N}}{t_n} + \sum_{i=-1}^{N-1}(-\mu)^i\diff{u_i}{t_n} - \mu\sum_s\diff{(\psi_s\varphi_s)}{t_n} = 0\\
      &\qquad\Longrightarrow\qquad
       (-\mu)^{N}\tilde{u}_{N} + \sum_{i=-1}^{N-1}(-\mu)^i u_i - \mu\sum_s \psi_s\varphi_s = a_n,
        \end{split}
      \end{equation}
  where $n\in\Z_+$ and $a_n$ is a time-independent function.
\end{itemize}
\end{theorem}
\begin{proof}
We already know that Lax operators \eqref{finiterestriction0} and
\eqref{finiterestriction1} generate consistent Lax hierarchies
\eqref{laxequation}. Thus, the right-hand side of
\eqref{laxequation} can be represented in the form of $L_{t_n}$.
Replacing $\delta$ with $-\mu^{-1}$ in \eqref{laxequation}, we have
\begin{equation}
    \left . L_{t_n} \right |_{\delta=-\mu^{-1}} =
    \left . [(L^n)_{\geqslant k},L] \right |_{\delta=-\mu^{-1}} = 0.
\end{equation}
Hence, the constraints \eqref{con1} and \eqref{con2} follow.
\end{proof}

The above theorem can be generalized to further restrictions. As a
consequence, the  constraints $\eqref{con1}$ or $\eqref{con2}$
with fixed common value of all $a_n$, are valid for the whole Lax
hierarchy \eqref{laxequation}.

\subsection{Recursion operators}

One of the characteristic features of integrable systems possessing
infinite-hierarchy of mutually commuting symmetries is the
existence of a recursion operator \cite{Olver,MB}. A recursion
operator of a given system, is an operator of such property that
when it acts on one symmetry of the system considered, it produces
another symmetry. G\"urses \emph{et al.} \cite{Gurses} 
presented a general and very efficient method of constructing recursion operators for Lax hierarchies.
Among others, the authors illustrated the method by applying it to
finite-field reductions of the KP hierarchy. In \cite{Blaszak} the
method was applied to the reductions of modified KP hierarchy as
well as to the lattice systems. Our further considerations are based
on the scheme from \cite{Gurses} and \cite{Blaszak}.

The recursion operator $\Phi$ has the following property:
\begin{equation*}
\Phi(L_{t_{n}})=L_{t_{n+N}},\qquad n\in\mathbb{Z}_{+},
\end{equation*}
and hence it allows reconstruction of the whole hierarchy
\eqref{laxequation} when applied to the first $(N-1)$ symmetries.
\begin{lemma}\label{recl}\hfill
 \begin{itemize}
  \item[(i)] The case $k=0$. Let the Lax operator be given in the general form  \eqref{finiterestriction0}.
Then, the recursion operator of the related Lax hierarchy can be
constructed solving
\begin{equation}\label{rec}
 L_{t_{n+N}} = L_{t_n}L + [R, L]
\end{equation}
with the remainder in the form
\begin{equation}\label{rem1}
 R = a_{N-1}\delta^{N-1} + \cdots + a_0 + \sum_s a_{-1,s}\delta^{-1}\varphi_s,
\end{equation}
where $N$ is the highest order of $L$.
  \item[(ii)] The case $k=1$. Similarly for the Lax operator \eqref{finiterestriction1}, the recursion operator can be constructed from \eqref{rec} with
\begin{equation}\label{rem2}
 R = a_N\delta^N + \cdots + a_0 + \sum_s a_{-1,s}\delta^{-1}\varphi_s.
\end{equation}
 \end{itemize}
\end{lemma}
\begin{proof}
Consider the case $k=0$. Then for \eqref{finiterestriction0} we
have
\begin{align*}
(L^\frac{n+N}{N})_{\geqslant 0} &=  ((L^\frac{n}{N})_{\geqslant 0}L)_{\geqslant 0} + ((L^\frac{n}{N})_{< 0}L)_{\geqslant 0}\\
    &= (L^\frac{n}{N})_{\geqslant 0}L - \sum_s[(L^\frac{n}{N})_{\geqslant 0}\psi_s]_0\delta^{-1}\varphi_s +
    ((L^\frac{n}{N})_{< 0}L)_{\geqslant 0}\\
    &= (L^\frac{n}{N})_{\geqslant 0}L + R,
\end{align*}
where  $[\sum_ia\delta^i]_0 = a_0$ and $R$ is given by
\eqref{rem1}. Similarly for $k=1$, we have
\begin{align*}
(L^\frac{n+N}{N})_{\geqslant 1} &=  ((L^\frac{n}{N})_{\geqslant 1}L)_{\geqslant 1} + ((L^\frac{n}{N})_{< 1}L)_{\geqslant 1}\\
    &= (L^\frac{n}{N})_{\geqslant 1}L -  [(L^\frac{n}{N})_{\geqslant 1}L]_0 -
    \sum_s[(L^\frac{n}{N})_{\geqslant 0}\psi_s]_0\delta^{-1}\varphi_s +  ((L^\frac{n}{N})_{< 1}L)_{\geqslant 1}\\
    &= (L^\frac{n}{N})_{\geqslant 1}L + R,
\end{align*}
where $R$ has the form \eqref{rem2}. Thus, in both cases
\eqref{rec} follows from \eqref{laxequation}. Hence we can extract
the recursion operator from \eqref{rec}.
\end{proof}

Note that in general, recursion operators on time scales are 
non-local. This means that they contain non-local terms with
$\Delta^{-1}$ being  formal inverse of $\Delta$ operator. However,
such recursion operators acting on an appropriate domain
produce only local hierarchies.

\section{Infinite-field integrable systems on time scales}

\subsection{Difference KP, $k=0$:}

 Consider the following infinite field Lax operator
\begin{equation}\label{kplax}
L=\delta+\tilde{u}_0+\sum_{i\geqslant 1}u_i\delta^{-i},
\end{equation}
which generates the Lax hierarchy \eqref{laxequation} as the
difference counterpart of the Kadomtsev-Petviashvili (KP)
hierarchy.

For $\displaystyle{(L)_{\geqslant 0}=\delta+\tilde{u}_0}$, the
first flow is given by
\begin{equation}\label{inf1ugeneral}
    \begin{split}
\frac{d\tilde{u}_0}{dt_1} &= \mu\Delta
u_1\\
\frac{du_i}{dt_1} &=
\sum_{k=0}^{i-1}(-1)^{k+1}u_{i-k}\sum_{j_1+j_2+\ldots+j_{k+1}=i}
(E^{-j_{k+1}}\Delta E^{-j_{k}}\Delta\ldots E^{-j_{2}}\Delta
E^{-j_{1}})\tilde{u}_0\\
&\qquad +\mu\Delta u_{i+1}+\Delta
u_{i}+u_{i}\tilde{u}_0\qquad\forall i>0,
    \end{split}
\end{equation}
where $j_\gamma >0$ for all $\gamma\geqslant 1$.

For $(L^2)_{\geqslant 0}=\delta^2+\xi\delta+\eta$, where
\begin{equation}
\xi:=E\tilde{u}_0+\tilde{u}_0\qquad \eta:=\Delta
\tilde{u}_0+\tilde{u}_0^2+u_1+Eu_1,
\end{equation}
one calculates the second flow
\begin{equation}\label{inf2ugeneral}
    \begin{split}
\frac{du_0}{dt_2} &= \mu\Delta(E+1)u_2+\mu\Delta(\Delta u_1+u_1\tilde{u}_0+u_1E^{-1}\tilde{u}_0)\\
\frac{du_i}{dt_2} &=
\sum_{k=-1}^{i-1}(-1)^{k+2}u_{i-k}\sum_{j_1+j_2+\ldots+j_{k+2}=i+1}
(E^{-j_{k+2}}\Delta E^{-j_{k+1}}\Delta\ldots E^{-j_{2}}\Delta E^{-j_{1}})\xi\\
&\qquad
+\sum_{k=0}^{i-1}(-1)^{k+1}u_{i-k}\sum_{j_1+j_2+\ldots+j_{k+1}=i}
(E^{-j_{k+1}}\Delta E^{-j_{k}}\Delta\ldots E^{-j_{2}}\Delta E^{-j_{1}})\eta\\
&\qquad +\Delta^2 u_i+(E\Delta+\Delta
E)u_{i+1}+\mu\Delta(E+1)u_{i+2}+ \xi(\Delta u_i+Eu_{i+1})+ \eta
u_i,
    \end{split}
\end{equation}
where $j_\gamma > 0$ for all $\gamma\geqslant1$. \\

The simplest case in $(2+1)$ dimensions: We
rewrite the first two members of the first flow by setting
$\tilde{u}_0=w$ and $t_1=y$ and the first member of the second flow by
setting $t_2=t$. Since $E$ and $\Delta$ do not commute, note that
in the calculations up to the last step, we use $E-1$ instead of
$\mu\Delta$, to avoid confusion.
\begin{eqnarray}
w_{y}&=&(E-1)u_1,\label{inf1w}\\
u_{1,y}&=&(E-1)u_2+\Delta u_1+
u_1(1-E^{-1})(w),\label{inf1u1y}\\
w_{t}&=&(E^2-1)u_2+(E-1)(\Delta u_1+
u_1w+u_1E^{-1}(w))\label{inf2ugeneralb}
\end{eqnarray}
Applying $E+1$ to \eqref{inf1u1y} from the left yields:
\begin{equation} (E^2-1)u_2=(E+1)u_{1,y}-(E+1)\Delta u_{1}-(E-1)u_{1}(1-E^{-1})w.\label{inf1wxx}  \end{equation}
Applying $(E-1)$ to \eqref{inf2ugeneralb} from the left and
substituting \eqref{inf1w} and \eqref{inf1wxx} into the new
derived equation we finally obtain the $(2+1)$-dimensional one-field
system of the form
\begin{equation}\mu\Delta w_t=(E+1)w_{yy}-2\Delta w_y+2\mu\Delta(ww_y).\label{inf2ugeneralc}\end{equation}
which does not have a continuous counterpart. For the case of
$\Time=h{\mathbb Z}$, one can show that \eqref{inf2ugeneralc} is
equivalent to the $(2+1)$-dimensional Toda lattice system.

The difference analogue of one-field continuous KP equation is too
complicated to write it down explicitly.
 \begin{remark}\label{remark1}
 Here we want to illustrate the behavior of $\tilde{u}_0$ in all symmetries of the difference KP hierarchy. Let
$\displaystyle{(L^n)_{< 0}=\sum_{i\geqslant1}
v^{(n)}_i\delta^{-i}}$, then by the right-hand of the Lax equation
\eqref{laxequation}, we obtain the first members of all flows
\begin{eqnarray}
\frac{d\tilde{u}_0}{dt_n}=\mu\Delta v^{(n)}_1.
\end{eqnarray}
Thus $\tilde{u}_0$  is time-independent for dense $x\in \Time$
since $\mu=0$. Hence when $\mathbb{T}=\mathbb{R}$, $\tilde{u}_{0}$
appears to be a constant.
\end{remark}

In $\Time=\Rm$ case, or in the continuous limit of some special
time scales, with $\tilde{u}_0=0$, the Lax operator \eqref{kplax}
turns out to be a Laurent series of pseudo-differential operators
\begin{equation}
L=\partial+\sum_{i\geqslant 1}u_{i}\partial^{-i}.
\end{equation}
Moreover, the first flow \eqref{inf1ugeneral} turns out to be
exactly the first flow of the  KP system
\begin{eqnarray}\frac{du_i}{dt_1}=u_{i,x},\qquad\forall i\geqslant
1\label{kp1}
\end{eqnarray}
while the second flow \eqref{inf2ugeneral} becomes exactly the
second flow of the  KP system
\begin{equation}\label{kp2}
\frac{du_i}{dt_2} =
(u_i)_{2x}+2(u_{i+1})_x+2\sum_{k=1}^{i-1}(-1)^{k+1}\binom{i-1}{k}
u_{i-k}(u_1)_{kx}\qquad\forall i\geqslant 1.
\end{equation}

\subsection{Difference mKP, $k=1$:}

Consider the Lax operator of the form
\begin{equation}\label{mkplax}
L=\tilde{u}_{-1}\delta+\sum_{i\geqslant 0}u_{i}\delta^{-i}
\end{equation}
which generates the difference counterpart of the modified
Kadomstsev-Petviashvili (mKP) hierarchy.

Then, $(L)_{\geqslant 1}=\tilde{u}_{-1}\delta$ implies the first
flow
\begin{equation}\label{inf3generala}
    \begin{split}
  \frac{d\tilde{u}_{-1}}{dt_1} &= \mu \tilde{u}_{-1}\Delta u_0\\
\frac{du_i}{dt_1} &=
\sum_{k=-1}^{i-1}(-1)^{k+2}u_{i-k}\sum_{j_1+j_2+\dots+j_{k+2}=i+1}(E^{-j_{k+2}}\Delta
E^{-j_{k+1}}\Delta\ldots E^{-j_{2}}\Delta E^{-j_{1}})\tilde{u}_{-1}\\
&\qquad + \tilde{u}_{-1}Eu_{i+1}+\tilde{u}_{-1}\Delta u_i\qquad
\forall i\geqslant 0,
    \end{split}
\end{equation}
where $j_\gamma>0$, $\gamma=1,2,\ldots,k+2$.

Next, for $(L^2)_{\geqslant 1}=\xi\delta^2+\eta\delta$, where
\begin{equation}
\xi:=\tilde{u}_{-1}E\tilde{u}_{-1},\qquad
\eta:=\tilde{u}_{-1}\Delta
\tilde{u}_{-1}+\tilde{u}_{-1}Eu_{0}+u_0\tilde{u}_{-1},
\end{equation}
we have the second flow as follows
\begin{equation}\label{inf4general}
    \begin{split}
  \frac{d\tilde{u}_{-1}}{dt_2} &= \xi(E\Delta u_0+E^2(u_1)) + \mu \tilde{u}_{-1}\Delta u_0^2
  - u_1E^{-1}\xi-\tilde{u}_{-1}^2\Delta u_0\\
\frac{du_i}{dt_2} &=
\sum_{k=-2}^{i-1}(-1)^{k+3}u_{i-k}\sum_{j_1+j_2+\ldots+j_{k+3}=i+2}
(E^{-j_{k+3}}\Delta E^{-j_{k+2}}\Delta\ldots\Delta E^{-j_{1}})\xi\\
&\qquad+\sum_{k=-1}^{i-1}(-1)^{k+2}u_{i-k}\sum_{j_1+j_2+\ldots+j_{k+2}=i+1}
(E^{-j_{k+2}}\Delta E^{-j_{k+1}}\Delta\ldots \Delta
E^{-j_{1}})\eta\\
&\qquad+\xi_2(\Delta^2 u_i+(E\Delta+\Delta
E)u_{i+1}+E^2u_{i+2})+\eta(\Delta u_{i}+Eu_{i+1}),
    \end{split}
\end{equation}
where $i\geqslant 0$ and $j_\gamma > 0$ for all $\gamma\geqslant
1$.

\begin{remark}\label{remark2} Similarly in order to   illustrate the behavior of $\tilde{u}_{-1}$
in all symmetries of the difference mKP hierarchy let us consider
$\displaystyle{(L^n)_{<1}=\sum_{i\geqslant
0}v^{(n)}_i\delta^{-i}}$. Then we obtain the first members of all
flows
\begin{eqnarray}
\frac{d\tilde{u}_{-1}}{dt_n}=\mu\tilde{u}_{-1}\Delta v^{(n)}_0,
\end{eqnarray} Thus $\tilde{u}_{-1}$  is time-independent for dense $x\in \Time$. Hence when $\mathbb{T}=\mathbb{R}$,
$\tilde{u}_{-1}$ appears to be a constant.
\end{remark}

In $\Time=\Rm$ case, or in the continuous limit of some special
time scales, with $\tilde{u}_{-1}=1$, the Lax operator
\eqref{mkplax} turns out to be the pseudo-differential operator
\begin{equation}
L=\partial+\sum_{i\geqslant 0}u_{i}\partial^{-i},
\end{equation}
Furthermore,  the system of equations \eqref{inf3generala} is
exactly the first flow of the mKP system
\begin{eqnarray}\frac{du_i}{dt_1}=u_{i,x},\qquad\forall
i\geqslant 0,\label{mkp1}
\end{eqnarray}
while the second flow \eqref{inf4general} turns out to be the
second flow of the mKP system
\begin{equation}\label{mkp2}
\begin{split}
\frac{du_i}{dt_2} &= (u_i)_{2x}+2(u_{i+1})_{x} + 2u_0(u_i)_x + 2u_0u_{i+1}\\
&\qquad+ 2\sum_{k=0}^{i}(-1)^{k+1}\binom{i}{k}
u_{i+1-k}(u_0)_{kx}\qquad\forall i\geqslant 0.
\end{split}
\end{equation}

\section{Finite-field integrable systems on time scales}

\subsection{Difference AKNS, $k=0$:}

Let the Lax operator \eqref{finiterestriction0} for $N=1$ and
$c_1=1$ is of the form
\begin{equation}\label{lax1}
L=\delta+\tilde{u}+\psi\delta^{-1}\varphi.
\end{equation}
The constraint \eqref{con1} between fields, with $a_n=0$, becomes
\begin{equation}\label{rel1}
    \tilde{u} = \mu \psi\varphi.
\end{equation}

For $(L)_{\geqslant 0}=\delta+\tilde{u}$, one finds the first flow
\begin{equation}\label{akns1}
    \begin{split}
      \frac{d\tilde{u}}{dt_1} &= \mu\Delta(\psi E^{-1}\varphi),\\
      \frac{d\psi}{dt_1} &= \tilde{u}\psi+\Delta\psi,\\
      \frac{d\varphi}{dt_1} &= -\tilde{u}\varphi+\Delta E^{-1}\varphi.
    \end{split}
\end{equation}
Eliminating field $\tilde{u}$ by \eqref{rel1} we have
\begin{equation}\label{akns11}
    \begin{split}
    \frac{d\psi}{dt_1}&=\mu\psi^2\varphi+\Delta\psi,\\
\frac{d\varphi}{dt_1}&=-\mu\varphi^2\psi+\Delta E^{-1}\varphi.
    \end{split}
\end{equation}

Next we calculate $\displaystyle{(L^2)_{\geqslant
0}=\delta^2+\xi\delta+\eta}$ where
\begin{equation}
\xi:=(E+1)\tilde{u},\quad \eta:=\Delta
\tilde{u}+\tilde{u}^2+\varphi E(\psi)+\psi E^{-1}(\varphi).
\end{equation}
Thus, the second flow takes the form
\begin{equation}\label{akns2}
    \begin{split}
    \frac{d\tilde{u}}{dt_2}&=\mu\Delta\brac{\Delta (\psi E^{-1}(\varphi))+\psi E^{-1}(\tilde{u}\varphi)+\tilde{u}\psi E^{-1}\varphi}-\mu\Delta(E+1)\psi E^{-1}\Delta E^{-1}(\varphi)\\
\frac{d\psi}{dt_2}&=\psi\eta+\xi\Delta\psi+\Delta^2\psi\\
\frac{d\varphi}{dt_2}&=-\varphi\eta+\Delta
E^{-1}(\xi\varphi)-(\Delta E^{-1})^2\varphi .
    \end{split}
\end{equation}
By the use of the constraint \eqref{rel1}, the second flow can be
written as
\begin{equation}\label{akns21}
    \begin{split}
\frac{d\psi}{dt_2}&=\psi(\Delta
\mu\psi\varphi+(\mu\psi\varphi)^2+\varphi E(\psi)+\psi
E^{-1}(\varphi))+
(E+1)\mu\psi\varphi\Delta\psi+\Delta^2\psi,\\
\frac{d\varphi}{dt_2}&=-\varphi(\Delta
\mu\psi\varphi+(\mu\psi\varphi)^2+\varphi E(\psi)+\psi
E^{-1}(\varphi))+\Delta E^{-1}(\varphi(E+1)\mu\psi\varphi)-(\Delta
E^{-1})^2\varphi .
    \end{split}
\end{equation}

In order to obtain the recursion operator one finds that for the
Lax operator \eqref{lax1} the appropriate reminder \eqref{rem1}
has the form
\begin{equation}
    R = \Delta^{-1}\bra{{\mu}^{-1}\tilde{u}_{t_n}} - \psi_{t_n}\delta^{-1}\varphi.
\end{equation}
Then, \eqref{rec} implies the following recursion formula as
\begin{equation}
    \pmatrx{\tilde{u}\\ \psi\\ \varphi}_{t_{n+1}} =
    \pmatrx{\tilde{u}-{\mu}^{-1} & \phi E & \psi E^{-1}\\ \psi+\psi\Delta^{-1}{\mu}^{-1} &
    \Delta+\tilde{u}+\psi\Delta^{-1}\varphi & \psi\Delta^{-1}\psi\\
    -\varphi\Delta^{-1}{\mu}^{-1} & -\varphi E\Delta^{-1}\varphi &
    \tilde{u} - \Delta E^{-1} - \varphi E\Delta^{-1}\psi}\pmatrx{\tilde{u}\\ \psi\\ \varphi}_{t_n}
\end{equation}
valid for isolated points $x\in\Time$, i.e. when $\mu\neq 0$. For
dense points one must use its reduction by constraint
\eqref{rel1}:
\begin{equation}\label{red}
    \pmatrx{\psi\\ \varphi}_{t_{n+1}} =
    \pmatrx{\Delta+\tilde{u}+\mu\psi\varphi + 2\psi\Delta^{-1}\varphi & \mu\psi^2+2\psi\Delta^{-1}\psi\\
    -\varphi(E+1)\Delta^{-1}\varphi &
    \tilde{u} - \Delta E^{-1} - \varphi(E+1)\Delta^{-1}\psi}\pmatrx{\psi\\ \varphi}_{t_n},
\end{equation}
where $\tilde{u}$ is given by \eqref{rel1}.

In $\Time=\Rm$ case, or in the continuous limit of some special
time scales, with the choice $\tilde{u}=0$, the Lax operator
\eqref{lax1} takes the form $L=\pr + \psi\pr^{-1}\varphi$. Then,
the continuous limits of \eqref{akns1} and \eqref{akns2}
respectively, imply that the first flow is the translational
symmetry
\begin{equation}
\begin{split}
 \frac{d\psi}{dt_1}&=\psi_x\\
 \frac{d\varphi}{dt_1}&= \varphi_x
\end{split}
\end{equation}
and the first non-trivial equation from the hierarchy is the AKNS
equation
\begin{equation}
    \begin{split}
    \frac{d\psi}{dt_2}&=\psi_{xx}+2\psi^2\varphi,\\
\frac{d\varphi}{dt_2}&=-\varphi_{xx}-2\varphi^2\psi .
    \end{split}
\end{equation}
For that special case the recursion formula \eqref{red} is of the
following form:
\begin{equation}
    \pmatrx{\psi\\ \varphi}_{t_{n+1}} =
    \pmatrx{\pr_x+2\psi\pr_x^{-1}\varphi & 2\psi\pr_x^{-1}\psi\\
    -2\varphi\pr_x^{-1}\varphi &
    - \pr_x - 2\varphi\pr_x^{-1}\psi}\pmatrx{\psi\\ \varphi}_{t_n}.
\end{equation}

\subsection{Difference Kaup-Broer, $k=1$:}

>From the admissible finite field restrictions
\eqref{finiterestriction1}, we consider the following simplest Lax
operator
\begin{equation}\label{lax2}
L=\tilde{u}\delta + v + \delta^{-1}w.
\end{equation}
The constraint \eqref{con2}, with $a_n=1$, implies
\begin{equation}\label{rel2}
    \tilde{u} = 1+\mu v-\mu^2w.
\end{equation}

Then, for $(L)_{\geqslant 1}=\tilde{u}\delta$, the first flow is
given as
\begin{equation}\label{1u}
\begin{split}
  \frac{d\tilde{u}}{dt_1}&=\mu \tilde{u}\Delta v,\\
\frac{dv}{dt_1}&=\tilde{u}\Delta v+\mu \Delta E^{-1}(\tilde{u}w),\\
\frac{dw}{dt_1}&=\Delta E^{-1}(\tilde{u}w).
\end{split}
\end{equation}
By the constraint \eqref{rel2} one can rewrite the first flow as
\begin{equation}
    \begin{split}\label{1u1}
     \frac{dv}{dt_1}&=(\mu v-\mu^2w)\Delta v+\mu \Delta E^{-1}(w(\mu v-\mu^2w)),\\
\frac{dw}{dt_1}&=\Delta E^{-1}\bra{\mu vw -\mu^2w^2}.
    \end{split}
\end{equation}

Next, we calculate $\displaystyle{(L^2)_{\geqslant
1}=\xi\delta^2+\eta\delta}$, where
\begin{equation}
\xi:=\tilde{u}E\tilde{u},\quad \eta:=\tilde{u}\Delta
\tilde{u}+\tilde{u}Ev+v\tilde{u},
\end{equation}
that yields the second flow
\begin{equation}\label{2u}
    \begin{split}
    \frac{d\tilde{u}}{dt_2}&=\mu \tilde{u}\Delta (E^{-1}+1)\tilde{u}w+\mu \tilde{u}\Delta v^2+\mu \tilde{u} \Delta (\tilde{u}\Delta v),\\
\frac{dv}{dt_2}&=\xi(\Delta^2v+\Delta w)+\mu\Delta
E^{-1}(w\eta)+E^{-1}\Delta E^{-1}(w\xi)+\eta \Delta v,\\
\frac{dw}{dt_2}&=-\Delta E^{-1}\Delta E^{-1}(w\xi)+\Delta
E^{-1}(w\eta).
    \end{split}
\end{equation}
One can rewrite the above system reducing it by the constraint,
but the final equation has complicated form.

For the Lax operator \eqref{lax2} the appropriate reminder
\eqref{rem2} is given by
\begin{equation}
    R = \tilde{u}\Delta^{-1}(\mu \tilde{u})^{-1}\tilde{u}_{t_n}\delta - v_{t_n} - \Delta^{-1}w_{t_n} .
\end{equation}
Hence, from \eqref{rec} we have the following, valid when $\mu\neq
0$, recursion formula
\begin{equation}
    \pmatrx{\tilde{u}\\ v\\ w}_{t_{n+1}} =
    \pmatrx{R_{\tilde{u}\tilde{u}} & \tilde{u}E & \mu \tilde{u}\\ R_{v\tilde{u}} & v + \tilde{u}\Delta & (1+E^{-1})\tilde{u}\\
    R_{w\tilde{u}} & w & -\Delta E^{-1}\tilde{u} + v - \mu w}
    \pmatrx{\tilde{u}\\ v\\ w}_{t_n},
\end{equation}
where
\begin{equation}
    \begin{split}
     R_{\tilde{u}\tilde{u}} &= E(v) -\mu^{-1}\tilde{u} + \mu \tilde{u}\Delta(v) \Delta^{-1}(\mu \tilde{u})^{-1} \\
    R_{v\tilde{u}} &= \Delta(v) + w + \tilde{u} \Delta(v) \Delta^{-1}(\mu \tilde{u})^{-1} + (1-E^{-1}) \tilde{u}w
    \Delta^{-1}(\mu \tilde{u})^{-1}\\
    R_{w\tilde{u}} &= \Delta E^{-1}\tilde{u}w\Delta^{-1}(\mu \tilde{u})^{-1}.
    \end{split}
\end{equation}
Its reduction by the constraint \eqref{rel2} is
\begin{equation}\label{red2}
    \pmatrx{v\\ w}_{t_{n+1}} =
    \pmatrx{v + \tilde{u}\Delta + R_{v\tilde{u}}\mu & (1+E^{-1})\tilde{u}-R_{v\tilde{u}}\mu^2\\
    w + R_{w\tilde{u}}\mu & -\Delta E^{-1}\tilde{u} + v - \mu w - R_{w\tilde{u}}\mu^2}
    \pmatrx{v\\ w}_{t_n},
\end{equation}
with $\tilde{u}$ given by \eqref{rel2}.

In the case of $\Time=\Rm$, or  in the continuous limit of some
special time scales, with the choice $\tilde{u}=1$, the Lax
operator \eqref{lax2} takes the form $L=\pr + v + \pr^{-1}w$. Then
the similar continuous analogue allows us to obtain the first flow
\begin{equation}
    \begin{split}
      \frac{dv}{dt_1}&=v_{x},\\
      \frac{dw}{dt_1}&=w_{x},
    \end{split}
\end{equation}
and the first non-trivial equation from the hierarchy is the
Kaup-Broer equation
\begin{equation}
    \begin{split}
      \frac{dv}{dt_2}&=v_{2x}+2w_{x}+2vv_{x},\\
\frac{dw}{dt_2}&=-w_{2x}+2(vw)_x.
    \end{split}
\end{equation}
For such special cases, the recursion formula \eqref{red2} turns
out to be
\begin{equation}
    \pmatrx{v\\ w}_{t_{n+1}} =
    \pmatrx{\pr_x + v + v_x\pr_x^{-1}& 2\\
    w + \pr_x w\pr_x^{-1} & -\pr_x + v}\pmatrx{v\\ w}_{t_n}.
\end{equation}

\section{Acknowledgments}

This work is partially  supported by the Scientific and Technical
Research Council of Turkey and MNiSW research grant N N202 404933.


\begin{thebibliography}{14}

\bibitem{MB} M. B\l aszak, {\it Multi-Hamiltonian Theory of Dynamical
Systems},  Texts and Monograhps in Physics (Springer-Verlag,
Berlin, 1998) 350pp.

\bibitem{frenkel}
E. Frenkel, {\it Deformations of the KdV hierarchy and related
soliton equations}, {\rm Int. Math. Res. Notices} {\bf 1996} 55
(1996)

\bibitem{Adler}
M. Adler, E. Horozov and P. van Moerbeke, {\it The Solution to the
q-kdv equation}, {\rm Phys. Lett. A. Notices} {\bf 242} 139 (1998)

\bibitem{G-G-S} M. G\"{u}rses, G. Sh. Guseinov, B. Silindir, {\it Integrable equations on time scales},
J. Math. Phys \textbf{46} (2005)113510

\bibitem{gelfand} I. M. Gelfand and L. A. Dickey, {\it Fractional powers of operators and Hamiltonian
systems},  {\rm Funct. Anal. Appl.} {\bf 10} 259-273 (1976)

\bibitem{bookdickey} L. A. Dickey, {\it Soliton equations and Hamiltonian systems},
{\rm Advenced studies in mathematical physics, volume 26 }, World
Scientific Publishing (2003)

\bibitem{bgss} M. B\l aszak, M. G\"urses, B. Silindir and B. M.
Szablikowski, {\it Integrable discrete systems on R and related
dispersionless systems}, {\rm arXiv: 0707.1084} (2007)

\bibitem{reyman} A. G. Reyman and M. A. Semenov-Tian-Shansky, {\it Family of Hamiltonian structures,
hierarchy of Hamiltonians and reduction for matrix first
order-differential operators}, {\rm Funkz. Analys. Priloz.} {\bf
14} 77-78 (1980)

\bibitem{semenov} M. A. Semenov-Tian-Shansky, {\it What is a classical
r-matrix?}, {\rm Funct. Anal. Appl.} {\bf 17} 259 (1983)

\bibitem{Gurses} M. G\"urses, A. Karasu and V.V. Sokolov, {\it On construction
of recursion operators from Lax representation}, J. Math. Phys.
{\bf 40} 6473-6490 (1999)

\bibitem{boh1} M. Bohner and A. Peterson, {\it Dynamic Equations on Time Scales:
An introduction with Applications}, Birkhauser, Boston (2001).

\bibitem{boh2} M. Bohner and A. Peterson, Editors, {\it Advances in Dynamic
Equations on Time Scales}, Birkhauser, Boston (2003).

\bibitem{ah} B. Aulbach and S. Hilger, {\it Linear Dynamic Process with
Inhomogeneous Time Scale}, in: {Nonlinear Dynamics and Quantum
Dynamical Systems} (Gaussing, 1990), {\rm Math. Res.,} {\bf 59},
Akademie Verlag, Berlin, 1990, pp.9-20.

\bibitem{hil} S. Hilger, {\it Analysis on measure chains--a unified approach to continuous and discrete calculus},
{\rm Results Math.}, {\bf 18}, 18-56 (1990).



\bibitem{aticiguseinov} F. M. Atici and G. Sh. Guseinov, {\it On Greens functions and positive solutions
for boundary value problems on time scales}, {\rm J. Comp. Appl.
Math.} {\bf 141} 75-99 (2002)

\bibitem{oevel} B. G. Konopelchenko  and W. Oevel, {\it An r-matrix approach to nonstandard classes of integrable equations},
 {\rm Publ. RIMS, Kyoto Univ.} {\bf 29} 581-666 (1993)

\bibitem{oevelsrt}W. Oevel and W. Strampp {\it Constrained KP hierarchy and bi-Hamiltonian structures},
 {\rm Commun. Math. Phys.}{\bf 157} 51 (1993)

\bibitem{Olver} P.J. Olver, {\it Applications of Lie Groups to Differential
Equations}, Springer New York 2000

\bibitem{Blaszak} M. B\l aszak, {\it On the construction of recursion operator and
algebra of symmetries for field and lattice systems}, Rep. Math.
Phys. {\bf 48} 27-38 (2001)


\end{thebibliography}
\end{document}